\documentstyle[11pt,fullpage,doublespace,epsf]{article}
 \DeclareMathSizes{11}{19}{13}{9}   

\makeatletter
\@addtoreset{equation}{section}
\makeatother

\begin{document}

   \title{ Riemann Hypothesis and Master Matrix for  FZZT Brane Partition Functions}

\author{Michael McGuigan\\Brookhaven National Laboratory\\Upton NY 11973\\mcguigan@bnl.gov}
\date{}
\maketitle

\begin{abstract}
We continue to investigate the physical interpretation of the Riemann zeta function as a FZZT brane partition function associated with a matrix/gravity correspondence begun in arxiv:0708.0645. We derive the master matrix of the $(2,1)$ minimal and $(3,1)$ minimal matrix model. We  use it's characteristic polynomial to understand why the zeros of the FZZT partition function, which is the Airy function, lie on the real axis. We also introduce an iterative procedure that can describe the Riemann $\Xi$ function as a deformed minimal model whose deformation parameters are related to a Konsevich integrand. Finally we discuss the relation of our work to other approaches to the Riemann $\Xi$ function including expansion in terms of Meixner-Pollaczek polynomials and Riemann-Hilbert problems.
\end{abstract}

\section{Introduction}

Recently it has been remarked that the gauge/gravity correspondence is more fascinating from a physics point of view than the Riemann hypothesis \cite{Horowitz:2006ct}. Ironically in our approach, perhaps they are manifestations of the same phenomena. 

In a previous paper we interpreted the Riemann $\Xi$ function as the FZZT partition function of a matrix model \cite{McGuigan:2007pr}. On the gauge or matrix side of the correspondence the FZZT brane partition function is the expectation value of the characteristic polynomial of a matrix $M$ weighted by $e^{-V(M)}$ with $V(M)$ the potential of a matrix model \cite{Fateev:2000ik}\cite{Teschner:2000md}\cite{Giusto:2004mt}\cite{Ellwood:2005nt}\cite{Hosomichi:2008th}. On the gravity side it is the exponentiated macroscopic loop \cite{Martinec:2004td}. The matrix model in question is of a complicated type, but seems related to the $(p,1)$ minimal matrix models solved in \cite{Hashimoto:2005bf} whose FZZT partition functions were the Airy function and generalized Airy functions \cite{Maldacena:2004sn}. These $(p,1)$ matrix models are special cases of the two matrix model \cite{Daul:1993bg}\cite{Kazakov:2004du}. Historical papers on matrix theories and the large $N$ limit are \cite{Wigner}\cite{Dyson:1972tm}\cite{'tHooft:1973jz}.

In this paper we construct master matrices for these theories and discuss the application to the zeros of the FZZT partition function. A master matrix is a particular large $N$ matrix whose characteristic polynomial yields the FZZT partition function in the large $N$ limit \cite{Gopakumar:1994iq}\cite{Gopakumar:1995bk}. This method allows us to understand in particular why the zeros of the Airy function are on the real axis. We also introduce an iterative method to define the $\Xi$ function that makes the relation to the matrix model clearer.

This paper is organized as follows. In section 2 we introduce the concept of the master matrix. We derive the master matrix of the $(2,1)$ minimal matrix model whose characteristic polynomial yields the FZZT partition function which is the Airy function. We also derive the master matrix of the $(3,1)$ minimal matrix model whose characteristic polynomial is the generalized Airy function. Interpreting the $\Xi$ function as a FZZT brane partition function, we discuss an iterative procedure which to leading order yields the deformed $(3,1)$  matrix model, to next order the deformed $(5,1)$  matrix model and so on. The deformation parameters are related to the expansion of the integrand of an integral representation of the $\Xi$ function which is similar to the integral representation of the generalized Airy functions. In that case the integrand is a Konsevich integrand which is cubic for the $(2,1)$ minimal model, quartic for the $(3,1)$ minimal model, sixth order for the $(5,1)$ minimal model etc. In section 3 we discuss the relation  of our approach to other approaches to the Riemann $\Xi$ function including the expansion of the $\Xi$ function in terms Meixner-Pollaczek polynomials and the definition of the Riemann $\Xi$ function as the solution to a Riemann-Hilbert problem. In section 4 we state the main conclusions of the paper.

\section{Master matrix}

\subsection{Master matrix of the $(2,1)$ minimal model}

The $(2,1)$ minimal model is defined by the partition function:
\[
\int dM dP e^{-V(M)+Tr(PM)}
\]
with:
\[
V(M) = \frac{1}{g}Tr(M^2)
\]
and $g$ is the coupling constant. In this paper we define a master matrix associated with the model as a matrix whose characteristic polynomial
is equal to the matrix integral:
\[
\int dM dP \det(M-zI)e^{-V(M)+Tr(PM)}
\]
which is the FZZT partition function.

Of particular interest is what happens as one takes the large $N$ limit, as the zeros of the FZZT partition function are related to the eigenvalues of the master matrix.
The master matrix for the $(2,1)$ minimal model is given by:
\begin{equation}
M = \sqrt {\frac{g}{2}} \left( {\begin{array}{*{20}c}
   0 & {\sqrt 1 } & 0 &  \ldots  & 0  \\
   {\sqrt 1 } & 0 & {\sqrt 2 } &  \ddots  &  \vdots   \\
   0 &  \ddots  &  \ddots  &  \ddots  & 0  \\
    \vdots  & 0 & {\sqrt {N - 2} } & 0 & {\sqrt {N - 1} }  \\
   0 &  \ldots  & 0 & {\sqrt {N - 1} } & 0  \\
\end{array}} \right)
\end{equation}
Which for $N=8$ is given by:
\[
\sqrt {g/2} \left( {\begin{array}{*{20}c}
   0 & {\sqrt 1 } & 0 & 0 & 0 & 0 & 0 & 0  \\
   {\sqrt 1 } & 0 & {\sqrt 2 } & 0 & 0 & 0 & 0 & 0  \\
   0 & {\sqrt 2 } & 0 & {\sqrt 3 } & 0 & 0 & 0 & 0  \\
   0 & 0 & {\sqrt 3 } & 0 & {\sqrt 4 } & 0 & 0 & 0  \\
   0 & 0 & 0 & {\sqrt 4 } & 0 & {\sqrt 5 } & 0 & 0  \\
   0 & 0 & 0 & 0 & {\sqrt 5 } & 0 & {\sqrt 6 } & 0  \\
   0 & 0 & 0 & 0 & 0 & {\sqrt 6 } & 0 & {\sqrt 7 }  \\
   0 & 0 & 0 & 0 & 0 & 0 & {\sqrt 7 } & 0  \\
\end{array}} \right)
\]
The FZZT partition function for the (2,1) minimal model was computed in \cite{Maldacena:2004sn} and is:
\[
\left( {\frac{g}{4}} \right)^{N/2} H_N (z/\sqrt g )
\]
This coincides with the characteristic polynomial of the master matrix. For the case $N=8$ this is:
\[
\frac{{105}}{{16}}g^4  - \frac{{105}}{2}g^3 z^2  + \frac{{105}}{2}g^2 z^4  - 14gz^6  + z^8 
\]
The master matrix (2.1) agrees with the master matrix of the Gaussian matrix model computed in \cite{Gopakumar:1994iq} which is has the same partition function as the $(2,1)$ minimal model after integration over $P$.

Because the master matrix is manifestly Hermitian it's eigenvalues are real. The large $N$ limit of FZZT partion function corresponds to  \cite{Hashimoto:2005bf}:
\[
\begin{array}{l}
 g \to \frac{1}{N} \\ 
 z \to  - 1 + \frac{1}{{N^{1/3} }}z \\ 
 \end{array}
\]
and leads to the Airy function $Ai(z)$.
This function is given by the contour integral:
\[
\Phi (z) = \int\limits_{C_0 } {\frac{{d\varphi }}{{2\pi i}}} e^{\varphi ^3 /3 - z\varphi } 
\]
with contour $C_0$ starting at infinity with argument $-\pi/3$ and ending at infinity with argument $\pi/3$. It has the series expansion:
\[
Ai(z) = \sum\limits_{n = 0}^\infty  {\frac{1}{{3^{2/3} \pi }}} \frac{{\Gamma ((n + 1)/3)}}{{n!}}\sin (2(n + 1)\pi /3)(3^{1/3} z)^n 
\]
The Airy function obeys the differential equation:
\[
Ai''(z) - zAi(z) = 0
\]
We plot the Airy function on the real line in Figure 1 and in the complex plane in Figure 2.
\begin{figure}[htbp]
  
   \centerline{\hbox{
   \epsfxsize=3.0in
   \epsffile{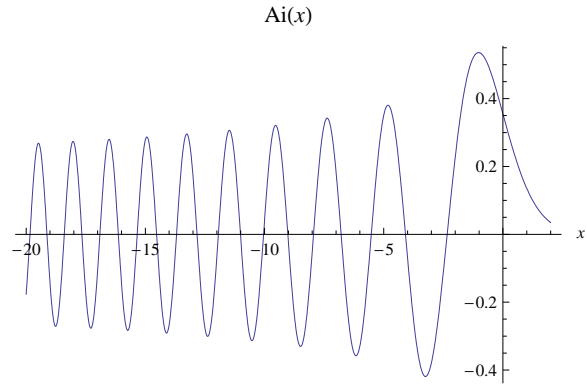}
     }
  }
 \caption{Plot of the of the Airy function on the real axis. }
  \label{fig1}
  
\end{figure}

\begin{figure}[htbp]
  
   \centerline{\hbox{
   \epsfxsize=3.0in
   \epsffile{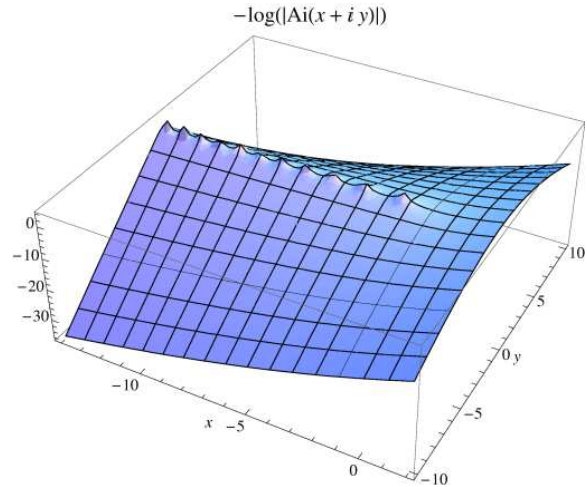}
     }
  }
 \caption{Plot of minus the logarithm of the magnitude of the  Airy function in the complex plane. The zeros are all located on the negative real axis. }
             
  \label{fig2}
  
\end{figure}
The Airy function  has all it's zeros on the real axis and this is a manifestation of the Hermitian nature of the master matrix in (2.1).

\subsection{ Master matrix of $(3,1)$ minimal model}

The $(3,1)$ minimal model is defined by the partition function  with matrix potential:
\[
V(M) = \frac{1}{g}(\frac{3}{2}Tr(M^2+\frac{1}{3}Tr(M^3))
\]
The master matrix of the $(3,1)$ minimal model is the matrix $M$ with nonzero components:
\[
M_{i,j} = (i-1)(i-2)\delta_{i,j+2} + 3(i-1)\delta_{i,j+1} + g\delta_{i+1,j}
\]
which is of the form:
\[
 M =  \left( {\begin{array}{*{20}c}
   0 & g & 0 &  \ldots  & 0  \\
   3 & 0 & g &  \ddots  &  \vdots   \\
   2 &  \ddots  &  \ddots  &  \ddots  & 0  \\
    \vdots  & (N-2)(N-3) & 3(N-2) & 0 & g  \\
   0 &  \ldots  & (N-1)(N-2) & 3(N-1) & 0  \\
\end{array}} \right)
\]
For $N=8$ this is given by:
\[
\left( {\begin{array}{*{20}c}
   0 & g & 0 & 0 & 0 & 0 & 0 & 0  \\
   3 & 0 & g & 0 & 0 & 0 & 0 & 0  \\
   2 & 6 & 0 & g & 0 & 0 & 0 & 0  \\
   0 & 6 & 9 & 0 & g & 0 & 0 & 0  \\
   0 & 0 & {12} & {12} & 0 & g & 0 & 0  \\
   0 & 0 & 0 & {20} & {15} & 0 & g & 0  \\
   0 & 0 & 0 & 0 & {30} & {18} & 0 & g  \\
   0 & 0 & 0 & 0 & 0 & {42} & {21} & 0  \\
\end{array}} \right)
\]
The characteristic polynomial of this master matrix for $g=1/N$ is given by:
\[
\frac{8085}{4096} - \frac{945 z}{256} - \frac{175 z^2}{8} + \frac{105 z^3}{16} + \frac{945 z^4}{32} - \frac{7 z^5}{4}
 - \frac{21 z^6}{2} + z^8
\]
and this corresponds to the FZZT partition function of the $(3,1)$ minimal model computed in \cite{Hashimoto:2005bf}.
\[
Q_N (z) = g^N (\partial _x )^N e^{ - \frac{1}{g}(\frac{1}{3}x^3  + \frac{3}{2}x^2  - xz)} |_{x = 0} 
\]
for $N=8$. The expression for $Q_N(z)$  can be written using the residue theorem as:
\[
Q_N (z) = ( - g)^N N!\frac{1}{{2\pi i}}\oint {\frac{{d\varphi }}{{\varphi ^{N + 1} }}} e^{ - V(\varphi ) - \varphi z} 
\]

After taking the large $N$ limit \cite{Hashimoto:2005bf}:
\[
\begin{array}{l}
 g \to \frac{1}{N} \\ 
 z \to  - 1 + \frac{1}{{N^{1/4} }}z \\ 
 \end{array}
\]
one obtains a generalized Airy function $\Phi(z)$ defined by the integral:
\[
\Phi (z) = \int\limits_{C_0 } {\frac{{d\varphi }}{{2\pi i}}} e^{\varphi ^4 /4 - z\varphi } 
\]
Similar integrals in the context of black holes were considered in \cite{Fidkowski:2003nf}. The generalized Airy function obeys the differential equation:
\[
\Phi '''(z) + z\Phi (x) = 0
\]
with solutions:
\begin{eqnarray*}
\Phi (z) &=&  A(_0 F_2 (\{ \} ,\{ \frac{1}{2},\frac{3}{4}\} , - \frac{{z^4 }}{{64}}))\\
& & {} + B(z^2 (_0 F_2 (\{ \} ,\{ \frac{5}{4},\frac{3}{2}\} , - \frac{{z^4 }}{{64}})))\\
& & {} + C(z(_0 F_2 (\{ \} ,\{ \frac{3}{4},\frac{5}{4}\} , - \frac{{z^4 }}{{64}})))
\end{eqnarray*}
for constants $A$, $B$, and $C$ where $pFq$ is a generalized hypergeometric function. The contour $C_0$ was chosen in \cite{Hashimoto:2005bf} so that one obtains a solution which is real for real $z$ and decays without oscillation for large positive $z$. 

Modifying the contour to be along the imaginary axis we can define a modified generalized Airy function $\Psi(z)$ by:
\[
\Psi (z) = \int_{ - \infty }^\infty  {e^{ - \frac{1}{4}\phi^4  + i\phi z} } d\phi
\]
with a series expansion  given by:
\[
\Psi (z) = \frac{1}{{\sqrt 2 }}\sum\limits_{k = 0}^\infty  {\frac{{( - 2)^k }}{{(2k)!}}} \Gamma (\frac{1}{4} + \frac{k}{2})z^{2k} 
\]
This modified generalized Airy function obeys the differential equation:
\[
\Psi '''(z) - z\Psi (x) = 0
\]
with solution:
\[
\Psi (z) = \frac{1}{{\sqrt 2 }}(\Gamma (\frac{1}{4}){}_0F_2 (\{ \} ,\{ \frac{1}{2},\frac{3}{4}\} ,\frac{{z^4 }}{{64}}) - z^2 \Gamma (\frac{3}{4}){}_0F_2 (\{ \} ,\{ \frac{5}{4},\frac{3}{2}\} ,\frac{{z^4 }}{{64}}))
\]
We plot the magnitude of this modified generalized Airy function $\Psi(z)$ on the real axis in Figure 3 and in the complex plane in Figure 4.
This function is even and as we shall see in the next section it has similar characteristics to the Riemann $\Xi$ function

\begin{figure}[htbp]
  
   \centerline{\hbox{
   \epsfxsize=3.0in
   \epsffile{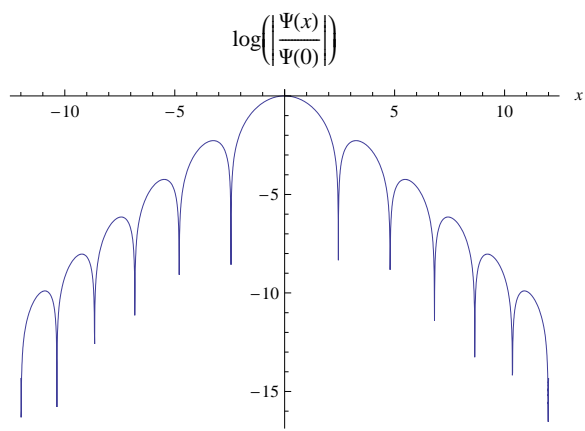}
     }
  }
 \caption{Plot of the of the logarithm of the magnitude of the $\Psi$ function on the real axis. The position of the zeros are given by the location of the spikes pointing down. }
  \label{fig3}
  
\end{figure}

\begin{figure}[htbp]
  
   \centerline{\hbox{
   \epsfxsize=3.0in
   \epsffile{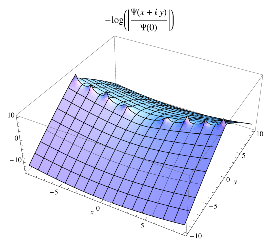}
     }
  }
 \caption{Plot of minus the logarithm of the magnitude of the $\Psi$ function in the complex plane. The zeros are symmetrically  located on the positive and negative real axis. The position of the zeros are given by the location of spikes pointing up.}
             
  \label{fig4}
  
\end{figure}

\subsection{Riemann $\Xi$ function}

The Riemann $\Xi$ function is defined by:
\[
\Xi (z ) = \zeta (i z  + \frac{1}{2})\Gamma (i\frac{{z }}{2} + \frac{1}{4})\pi ^{ - 1/4} \pi ^{ - iz /2} (
- \frac{{z ^2 }}{2} - \frac{1}{8})
\]
It is even and can be expressed as an integral along the imaginary axis as:
\[
\Xi (z) = \int_{ - \infty }^\infty  {e^{ - U(\phi)  + i\phi z} } d\phi
\]
where:
\[
U(\phi) = -\log({\sum\limits_{k = 1}^\infty  {(\pi ^2 k^4 }
e^{2\phi }  - \frac{3}{2}\pi k^2 e^{\phi } )e^{ - \pi k^2 e^\phi }})
\]
This function plays the same role for the $\Xi$ function as the Konsevich potential $\phi^3/3$ plays for the Airy function and $\phi^4/4$ for the $\Phi$ function \cite{Kontsevich:1992ti}\cite{Kharchev:1992dj}\cite{Gaiotto:2003yb}. We plot the function $U(\phi)$ in Figure 5. It is even and this leads to the fact that the $\Xi$ function is even.
\begin{figure}[htbp]
  
   \centerline{\hbox{
   \epsfxsize=3.0in
   \epsffile{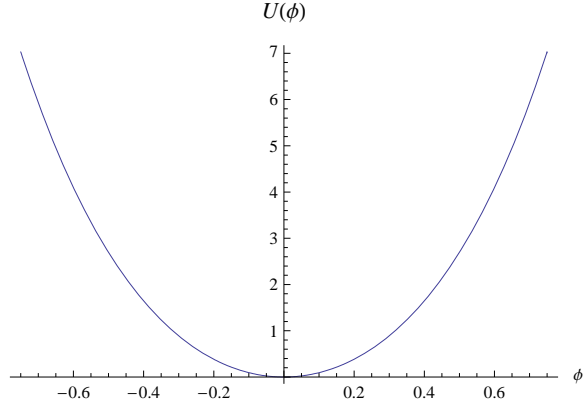}
     }
  }
 \caption{Plot of  the function $U(\phi)$. The relation $U(\phi) = U(-\phi)$ leads to the fact that the $\Xi$ function is even.}

  \label{fig5}
  
\end{figure}
For small $\phi$ one can develop an expansion:
\begin{equation}
U(\phi)=9.36345 \phi^2 + 5.95896 \phi^4 - 2.15104 \phi^6 + O(\phi^8)
\end{equation}
which is probably why the $(3,1)$ minimal model modified FZZT partition function shares some of the characteristics of the $\Xi$ function.

The $\Xi$ function itself can be expanded as \cite{Edwards}:
\[
\Xi (z )  = \sum\limits_{n = 0}^\infty  {a_{2n} \frac{{( - 1)^n }}{{\left(
{2n} \right)!}}} z ^{2n}
\]
where
\[
a_{2n}  = 4\int\limits_1^\infty  {d\ell (\ell ^{ - 1/4} f(\ell )(\frac{1}{2}\log \ell )^{2n} } )
\]
and
\[
f(\ell ) = \sum\limits_{q = 1}^\infty  {(q^4 \pi ^2 \ell }  - \frac{3}{2}q^2 \pi )\ell ^{1/2} e^{ - q^2 \pi \ell }
\]
Thus like the $\Psi$ function one can think of the $\Xi(z)$ function as an infinite order polynomial expanded in even powers of $z$. We plot the magnitude of the $\Xi$ function on the real axis in Figure 6 and in the complex plane in Figure 7. The Riemann hypothesis is equivalent to the statement that the zeros of the $\Xi$ function lie on the real axis.

\begin{figure}[htbp]
  
   \centerline{\hbox{
   \epsfxsize=3.0in
   \epsffile{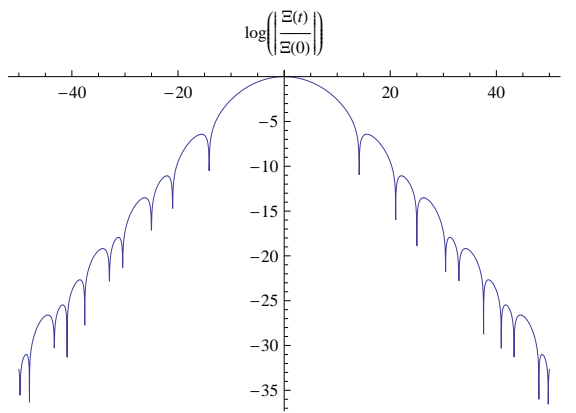}
     }
  }
 \caption{Plot of  the logarithm of the magnitude of the $\Xi$ function on the real axis. The zeros are symmetrically  located on the positive and negative real axis. The position of the zeros are given by the location of spikes pointing down.}
             
  \label{fig6}
  
\end{figure}

\begin{figure}[htbp]
  
   \centerline{\hbox{
   \epsfxsize=3.0in
   \epsffile{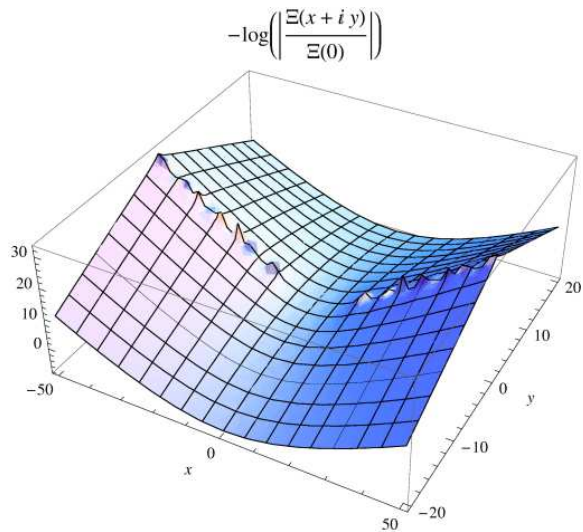}
     }
  }
 \caption{Plot of minus the logarithm of the magnitude of the $\Xi$ function in the complex plane. The zeros are symmetrically  located on the positive and negative real axis. The position of the zeros are given by the location of spikes pointing up.}
             
  \label{fig7}
  
\end{figure}

The Riemann $\Xi$ function does not obey a finite order differential equation. Nevertheless keeping the first two terms in the expansion for $U(\phi)$ one can derive the following approximate equation for small $z$:
\begin{equation}
4(5.95896)\Xi(z)''' -2(9.36345)\Xi(z)' -z\Xi(z) \approx  0
\end{equation}
This can be seen to be related to the generalized Airy equation with a deformed matrix potential.

Reversing the process of the previous subsection which was (1) matrix potential (2) master matrix (3) Orthogonal polynomial (4) contour integral (4) generalized Airy function (5)generalized Airy differential equation, one can attempt to reconstruct a master matrix.

Rescaling the argument of $\Xi(z)$ we define:
\[
\Xi_{*}(z) = \Xi(\sqrt{2}(5.95896))^{1/4}z)
\]
So that one has the following approximate equation for small $z$:
\[
\Xi_{*}(z)''' -s_1\Xi_{*}(z)' - z\Xi_{*}(z) \approx  0
\]
where:
\begin{equation}
s_1=\frac{9.36345}{\sqrt{5.95896}}
\end{equation}
This appears related to the deformed $(3,1)$ minimal model discussed in \cite{Hashimoto:2005bf} with deformation parameter $s_1$, in the same way that the function $\Psi$ was related to $\Phi$ in the undeformed $(3,1)$ model.

The solution to the equation for $\Xi*$ is denoted by $\Psi(z,s_1)$ and is:
 \[
\Psi (z,s_1) = \int_{ - \infty }^\infty  {e^{ - \frac{1}{4}\phi^4 -\frac{1}{2}s_1\phi^2 + i\phi z} } d\phi
\]
We plot this function in Figure 8 and note the qualitative similarity to the $\Xi(z)$ function for small $z$.
\begin{figure}[htbp]
  
   \centerline{\hbox{
   \epsfxsize=3.0in
   \epsffile{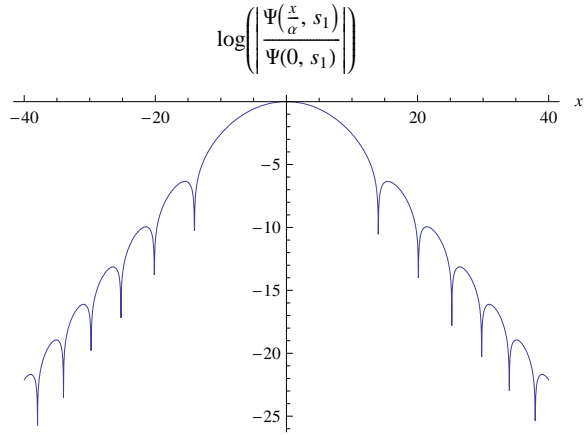}
     }
  }
 \caption{
Plot of  the logarithm of the magnitude of the $\Psi(x/\alpha,s_1)$ function on the real axis for $s_1= \frac{(9.36345)}{(5.95896))^{1/2}}$ and $\alpha = \sqrt{2} (5.95896)^{1/4}$. The zeros are symmetrically  located on the positive and negative real axis. The position of the zeros are given by the location of spikes pointing down.}
             
  \label{fig8}
  
\end{figure}

One can improve the approximate equation (2.3) by including higher order terms in the $\phi$ expansion of $U(\phi)$. Keeping terms up to $\phi^6$ in (2.2) one obtains the approximate differential equation :

\[
6(2.15104)\Xi'''''(z) + 4(5.95896)\Xi(z)''' -2(9.36345)\Xi(z)' -z\Xi(z) \approx  0
\]
Now rescaling can put the equation in the form:

\[
\Xi_{**}(z)''''' + s_3\Xi_{**}(z)''' -s_1\Xi_{**}(z)' -z\Xi_{**}(z) \approx  0
\]
with deformation parameters $s_1$ and $s_3$. This equation appears  related to the deformation of the $(5,1)$ minimal model of \cite{Hashimoto:2005bf}.

Finally we can define a function $\Phi(z,s1)$ as the solution to:
\begin{equation}
\Phi(z,s_1)''' -s_1\Phi(z,s_1)' + z\Phi(z,s_1) =  0
\end{equation}
which is real on the real axis and decays non-oscillatory for large positive $y$.
Using the results of \cite{Hashimoto:2005bf} $\Phi(z,s_1)$ is the FZZT partition function associated with the matrix potential:
\[
V(x) = \frac{N}{{1 + \frac{{s_1 }}{{\sqrt N }}}}(3( - 1 + \frac{x}{{N^{1/4} }}) + \frac{3}{2}( - 1 + \frac{x}{{N^{1/4} }})^2  + \frac{1}{3}( - 1 + \frac{x}{{N^{1/4} }})^3  + \frac{{s_1 }}{{\sqrt N }}( - 1 + \frac{x}{{N^{1/4} }}))
\]
After rescaling and shifting the point of origin of the potential one can define polynomials for the matrix model deformed by the parameter $s_1$ through:
\[
Q_N (z,s_1 ) = \left( {\frac{{1 + \frac{{s_1 }}{{\sqrt N }}}}{N}} \right)^N \partial _x^N (exp( - \frac{N}{{1 + \frac{{s_1 }}{{\sqrt N }}}}(3x + \frac{3}{2}x^2  + \frac{1}{3}x^3  + \frac{{s_1 }}{{\sqrt N }}x - xz))|_{x = 0} 
\]
The master matrix which has this as characteristic polynomial is a simple rescaling of the coupling constant of the master matrix of the $(3,1)$ minimal model and is given by:
\[
M_{i,j} = (i-1)(i-2)\delta_{i,j+2} + 3(i-1)\delta_{i,j+1} 
+ \frac{1}{N}(1+\frac{s_1}{\sqrt{N}})\delta_{i+1,j}
\] 
This master matrix can develop complex eigenvalues for large enough $N$ and $s_1$. In particular for $N \ge 34$ and $s_1$ given by (2.4) the eigenvalues are complex. However the function $\Psi(z,s_1)$ obtained from changing the sign of $z$ in the third term in (2.5) is very different from $\Phi(z,s_1)$ in this respect. It would be of interest to determine the master matrix associated with $\Psi(z,s_1)$ and it's corrections for terms involving $s_3$ and higher, which should in principle converge to the Riemann $\Xi$ function.

\section{Other related approaches}

In this section we compare our approach with other approaches to the Riemann $\Xi$ function.

\subsection{Expansion in Meixner-Pollaczek polynomials}

In \cite{Kuznetsov} the $\Xi$ function was expanded in Meixner-Pollaczek polynomials. These can be expressed as :
\begin{equation}
p_n (z) = n!\frac{{(2n + 1)!!}}{{(2n)!!}}i^n{}_2F_1 ( - n,\frac{3}{4} + i\frac{1}{2}z;\frac{3}{2};2)
\end{equation}
These polynomials are the characteristic polynomial of a matrix with nonzero components:
\[
M_{i,j}= i(i+\frac{1}{2})\delta_{i,j+1} + \delta_{i+1,j}
\]

For $N=8$ this matrix is given by:
\begin{equation}
\left( {\begin{array}{*{20}c}
   0 & 1 & 0 & 0 & 0 & 0 & 0 & 0  \\
   {3/2} & 0 & 1 & 0 & 0 & 0 & 0 & 0  \\
   0 & 5 & 0 & 1 & 0 & 0 & 0 & 0  \\
   0 & 0 & {21/2} & 0 & 1 & 0 & 0 & 0  \\
   0 & 0 & 0 & {18} & 0 & 1 & 0 & 0  \\
   0 & 0 & 0 & 0 & {55/2} & 0 & 1 & 0  \\
   0 & 0 & 0 & 0 & 0 & {36} & 0 & 1  \\
   0 & 0 & 0 & 0 & 0 & 0 & {105/2} & 0  \\
\end{array}} \right)
\end{equation}
The characteristic polynomial of this matrix is:
\[
\frac{{363825}}{{16}} - \frac{{74247}}{2}z^2  + \frac{{10493}}{2}z^4  - 154z^6  + z^8 
\]
which agrees with (3.1) for $N=8$.

The expansion of the $\Xi$ function with an exponential factor can be expanded in terms of the Meixner-Pollaczek polynomials as \cite{Kuznetsov}:
\[
\Xi(z)e^{-\pi z/4}  = \sum\limits_{n = 0}^{\infty} {b_n p_n(z) } 
\]
Terminating this series at $N$ one can write this expansion as the characteristic polynomial of a $N\times N$ matrix. For $N=8$ this is given by:
\[
\left( {\begin{array}{*{20}c}
   0 & 1 & 0 & 0 & 0 & 0 & 0 & 0  \\
   {3/2} & 0 & 1 & 0 & 0 & 0 & 0 & 0  \\
   0 & 5 & 0 & 1 & 0 & 0 & 0 & 0  \\
   0 & 0 & {21/2} & 0 & 1 & 0 & 0 & 0  \\
   0 & 0 & 0 & {18} & 0 & 1 & 0 & 0  \\
   0 & 0 & 0 & 0 & {55/2} & 0 & 1 & 0  \\
   0 & 0 & 0 & 0 & 0 & {36} & 0 & 1  \\
   {b_0 /b_8 } & {b_1 /b_8 } & {b_2 /b_8 } & {b_3 /b_8 } & {b_4 /b_8 } & {b_5 /b_8 } & {105/2 + b_6 /b_8 } & {b_7 /b_8 }  \\
\end{array}} \right)
\]
When $b_n$ are taken to zero this reproduces the matrix (3.2).

The coefficients $b_n$ are linearly related to the integrals \cite{Kuznetsov}:
\[
I_n  = \int_0^\infty  {\frac{{\sin (\frac{{y^2 }}{2} + \frac{\pi }{8})}}{{e^{2\sqrt \pi  y}  + 1}} y^{2n + 1} dy}
\]
Unlike the $a_n$ of the previous section there are closed form expressions for these integrals \cite{Kuznetsov}.
There is also some indication that there is some numerical advantage to the computation of $b_n$
at large $n$ using the asymptotic expansion of the analytic expression. If one thinks of the origin of the $\Xi$ function as coming from a quantum mechanical system then the expansion of the function in terms of different polynomials is similar to different choices of basis functions for the quantum description. Although physically there is no difference for quantum physics from the choice of basis, numerically there is some advantage if overlap integrals can be performed analytically.

\subsection{Riemann-Hilbert formulation}

Besides the large $N$ approach to matrix models one can develop a Riemann-Hilbert formulation of these theories \cite{Its}\cite{Konig}\cite{Brezin}\cite{Kitaev}\cite{Strahov}\cite{Gangardt}. In \cite{Its} it is shown that  one can use contour integrals to produce Riemann-Hilbert representations of special functions like the Airy function and Riemann zeta function. The jump matrix is an upper triangular matrix of the form:
\[
G(\varphi ,z) = \left( {\begin{array}{*{20}c}
   1 & {2\pi ig(\varphi ,z)}  \\
   0 & 1  \\
\end{array}} \right)
\]
and the function of interest is given by the contour integral:
\[
I(z) = \int_C {g(\varphi ,z)d\varphi } 
\]
or alternatively as the unique solution to the Riemann-Hilbert problem determined by the pair $(G,C)$.
For the Airy function $g(\phi,z)$ is given by \[
g(\varphi ,z) = e^{i\varphi ^3 /3 + i\varphi z} 
\] 
while for the Riemann zeta function it is given by
\[
g(\varphi ,z) = e^{ - U(\varphi ) + i\varphi z} 
\]
Thus in the Riemann-Hilbert formulation the relation between the jump matrix for the Airy function and the zeta function is the same as in the previous section, namely the replacement of $i\frac{\phi^3}{3}$ with $-U(\phi)$.

\section{Conclusion}

We have discussed the master matrix formulation of FZZT partition functions. We derived the master matrix associated with the $(2,1)$ minimal model and found agreement with the results of \cite{Gopakumar:1994iq}. The characteristic polynomial of the master matrix was the FZZT partition function of the $(2,1)$ model which is the Airy function after taking the large $N$ limit. We also derived the master matrix of he $(3,1)$ minimal model and related it's characteristic polynomial with the $(3,1)$ FZZT partition function which is a generalized Airy function $\Phi(z)$. In both cases the zeros of the FZZT partition function were on the real axis. The extension to the general $(p,1)$ minimal model should be straightforward. Interpreting the Riemann $\Xi$ function as a FZZT partition function we developed a controlled expansion of the function in terms of parameters $s_1,s_3,\ldots$ of a Konsevich type potential $U(\phi)$. We showed how this is related to the FZZT partition function of a minimal model with deformation parameters given by $s_i$. For order $s_1$ one uses a deformed $(3,1)$ matrix model to describe the theory. To order $s_3$ one uses a deformed $(5,1)$ model etc. This procedure should converge to the Riemann $\Xi$ function as one includes higher and higher order terms in the expansion of $U(\phi)$. More work relating the master matrix of the FZZT partition functions $\Phi(z,s_1,s_3,\ldots)$ to the function $\Psi(z,s_1,s_3,\ldots)$ is needed to obtain quantitative and qualitative insight into the arrangement of the zeros of the $\Psi(z,s_1,s_3,\ldots)$ and $\Xi(z)$in the complex plane. However a comparison between Figure 6 and Figure 8 is encouraging. Finally we compared our approach to other approaches to the Riemann $\Xi$ function which involve large $N$ matrices associated with expansion in Meixner-Pollaczek polynomials and the definition of the Airy and $\Xi$ functions as the solution of a Riemann-Hilbert problem.

\section*{Acknowledgements}
We thank David Shih and Simeon Hellerman for useful discussions and the Simons Conference at Stony Brook for hospitality. This manuscript has been authored in part by Brookhaven Science Associates, LLC, under Contract No. DE-AC02-98CH10886 with the U.S. Department of Energy.


\begin{thebibliography}{100}


\bibitem{Horowitz:2006ct}
  G.~T.~Horowitz and J.~Polchinski,
  ``Gauge / gravity duality,''
  arXiv:gr-qc/0602037.

\bibitem{McGuigan:2007pr}
  M.~McGuigan,
  ``Riemann Hypothesis, Matrix/Gravity Correspondence and FZZT Brane Partition
  Functions,''
  arXiv:0708.0645 [math-ph].

\bibitem{Fateev:2000ik}
  V.~Fateev, A.~B.~Zamolodchikov and A.~B.~Zamolodchikov,
  ``Boundary Liouville field theory. I: Boundary state and boundary  two-point
  function,''
  arXiv:hep-th/0001012.
\bibitem{Teschner:2000md}
  J.~Teschner,
  ``Remarks on Liouville theory with boundary,''
  arXiv:hep-th/0009138.
\bibitem{Giusto:2004mt}
  S.~Giusto and C.~Imbimbo,
  ``The Kontsevich connection on the moduli space of FZZT Liouville branes,''
  Nucl.\ Phys.\  B {\bf 704}, 181 (2005)
  [arXiv:hep-th/0408216].


\bibitem{Ellwood:2005nt}
  I.~Ellwood and A.~Hashimoto,
  ``Open / closed duality for FZZT branes in c = 1,''
  JHEP {\bf 0602}, 002 (2006)
  [arXiv:hep-th/0512217].

\bibitem{Hosomichi:2008th}
  K.~Hosomichi,
  ``Minimal Open Strings,''
  arXiv:0804.4721 [hep-th].

\bibitem{Martinec:2004td}
  E.~J.~Martinec,
  ``Matrix models and 2D string theory,''
  arXiv:hep-th/0410136.


\bibitem{Hashimoto:2005bf}
  A.~Hashimoto, M.~x.~Huang, A.~Klemm and D.~Shih,
  ``Open / closed string duality for topological gravity with matter,''
  JHEP {\bf 0505}, 007 (2005)
  [arXiv:hep-th/0501141].


\bibitem{Maldacena:2004sn}
  J.~M.~Maldacena, G.~W.~Moore, N.~Seiberg and D.~Shih,
  ``Exact vs. semiclassical target space of the minimal string,''
  JHEP {\bf 0410}, 020 (2004)
  [arXiv:hep-th/0408039].


\bibitem{Daul:1993bg}
  J.~M.~Daul, V.~A.~Kazakov and I.~K.~Kostov,
  ``Rational theories of 2-D gravity from the two matrix model,''
  Nucl.\ Phys.\  B {\bf 409}, 311 (1993)
  [arXiv:hep-th/9303093].

\bibitem{Kazakov:2004du}
  V.~A.~Kazakov and I.~K.~Kostov,
  ``Instantons in non-critical strings from the two-matrix model,''
  arXiv:hep-th/0403152.

\bibitem{Wigner}
Eugene Paul Wigner "On the statistical distribution of the widths and spacings of nuclear resonance levels". Proc. Cambr. Philos. Soc. 47: 790 (1951).

\bibitem{Dyson:1972tm}
  F.~J.~Dyson,
  ``A class of matrix ensembles,''
  J.\ Math.\ Phys.\  {\bf 13}, 90 (1972).

\bibitem{'tHooft:1973jz}
  G.~'t Hooft,
  ``A planar diagram theory for strong interactions,''
  Nucl.\ Phys.\  B {\bf 72}, 461 (1974).

\bibitem{Gopakumar:1994iq}
  R.~Gopakumar and D.~J.~Gross,
  ``Mastering the master field,''
  Nucl.\ Phys.\  B {\bf 451}, 379 (1995)
  [arXiv:hep-th/9411021].


\bibitem{Gopakumar:1995bk}
  R.~Gopakumar,
  ``The master field revisited,''
  Nucl.\ Phys.\ Proc.\ Suppl.\  {\bf 45B}, 244 (1996).

\bibitem{Fidkowski:2003nf}
  L.~Fidkowski, V.~Hubeny, M.~Kleban and S.~Shenker,
  ``The black hole singularity in AdS/CFT,''
  JHEP {\bf 0402}, 014 (2004)
  [arXiv:hep-th/0306170].




\bibitem{Kontsevich:1992ti}
  M.~Kontsevich,
  ``Intersection theory on the moduli space of curves and the matrix Airy
  function,''
  Commun.\ Math.\ Phys.\  {\bf 147}, 1 (1992).

\bibitem{Kharchev:1992dj}
  S.~Kharchev, A.~Marshakov, A.~Mironov and A.~Morozov,
  ``Generalized Kontsevich model versus Toda hierarchy and discrete matrix
  models,''
  Nucl.\ Phys.\  B {\bf 397}, 339 (1993)
  [arXiv:hep-th/9203043].

\bibitem{Gaiotto:2003yb}
  D.~Gaiotto and L.~Rastelli,
  ``A paradigm of open/closed duality: Liouville D-branes and the  Kontsevich
  model,''
  JHEP {\bf 0507}, 053 (2005)
  [arXiv:hep-th/0312196].


\bibitem{Edwards}
H.~M.~Edwards, "Riemann Zeta Function", Dover (1974).

\bibitem{Kuznetsov}
A.~Kuznetsov, ``Expansion of the Riemann $\Xi$ function in Meixner-Pollaczek polynomials'', (2006).\\
http://www.unbsj.ca/sase/math/faculty/akuznets/publications.html

\bibitem{Its}
A.~Its, "The Riemann-Hilbert problem and integrable sysytems", Notices of the AMS, 1389 (2003).

\bibitem{Konig}
W.~Konig, "Orthogonal polynomial ensembles in probability theory",  
Probability Surveys, Vol.2, 385 (2005).

\bibitem{Brezin}
E.~Brezin, S. ~Hikami, "Characteristic polynomials in random matrix theory", Commun.Math.Phys.214,111 (2000).

\bibitem{Kitaev}
A.~Kitaev, "Special functions of the isomonodromy type", Acta Applicandae Mathematicae 64, 1 (2000).

\bibitem{Strahov}
E.~Strahov, Y. Fyodorov, "Universal results for correlations of characteristic polynomials: Riemann-Hilbert approach", math-ph/0210010 (2002).

\bibitem{Gangardt}
D.~Gangardt, "Second quantization approach to characteristic polynomials in RMT",
nlin.CD/0011014 (2000).


\end{thebibliography}
\end{document}